\documentstyle[fleqn,12pt]{article}
 
\newtheorem{example}{Example}
\newtheorem{thm}{Theorem}
\newtheorem{lemma}[thm]{Lemma}
\newtheorem{cor}[thm]{Corollary}
\newtheorem{prop}[thm]{Proposition}

\newtheorem{composite-transformation}{Composite Transformation}

\newenvironment{proof}{
\setlength{\parindent}{0mm} \setlength{\parskip}{5mm}
{\bf Proof:} }{$\Box$}

\newcommand{\eqns}{\[ \begin{array}{l}}
\newcommand{\eqnsend}{\end{array} \]}

\newsavebox{\headering}

\newcommand{\ignore}[1]{}

\newcounter{lblno}

\newcounter{stepno}

\makeatletter
\renewcommand\section{\@startsection {section}{1}{\z@}{-0.5ex plus -0.2ex minus
 -.2ex}{0.5ex plus .2ex}{\Large\bf}}
\renewcommand\subsection{\@startsection{subsection}{2}{\z@}{-0.3ex plus -0.2ex minus
 -.2ex}{0.3ex plus .2ex}{\large\bf}}
\renewcommand\subsubsection{\@startsection{subsubsection}{3}{\z@}{-0.3ex plus
 -0.2ex minus -.2ex}{0.3ex plus .2ex}{\normalsize\bf}}
\makeatother

\newcommand{\ov}[1]{\mbox{$\overline{#1}$}}
\newcommand{\comment}[1]{}
\newcommand{\finish}[1]{}
 
 \title{A lemma on closures and its application to modularity in logic programming semantics}
 
 \author{Michael J. Maher \\
 	\emph{Reasoning Research Institute\footnote{
	This is a technical report of the Reasoning Research Institute.}} \\
	\emph{Canberra, Australia}\\
	Email:  \texttt{michael.maher@reasoning.org.au}
	}

 \date{Written 1991, released 2020}

\begin{document}

\maketitle

\def	\comp	{\circ}
\def	\lsb	{[\![}
\def	\rsb	{]\!]}
\def	\lan	{\langle}
\def	\ran	{\rangle}
\def	\cC	{{\cal C}}
 
\begin{abstract}
This note points out a lemma on closures of monotonic increasing functions
and shows how it is applicable to decomposition and modularity
for semantics defined as the least fixedpoint of some monotonic function.
In particular it applies to numerous semantics of logic programs.
An appendix addresses the fixedpoints of (possibly non-monotonic) functions
that are sandwiched between functions with the same fixedpoints.
\end{abstract}

{\bf  Note:}
This is a cleaned up version of a draft, probably begun in 1990, and last revised in 1991
(before the cleaning-up).
It has been cleaned up by:
 completing references (some were incomplete or the publication had not yet appeared),
 deleting notes to self,
and
 adding a little structure (including section headings).
The note is lacking introduction and motivational text, and a more detailed discussion of related work.
The appendix dates from some later time. 
  
\section{Preliminaries}
We assume a fixed domain of computation.
It can be any constraint domain,
but if it is a domain of finite (or rational or infinite or...) trees
then the set of function symbols is fixed in advance,
and is independent of the program(s).
 
A {\em module} $P$ is a pair $\lan R, S\ran$
where $R$ is a set of rules and $S$ is a set of ground atoms,
the set of ground atoms whose truth value $R$ defines.
To avoid some difficulties with arbitrary use of this definition,
in this paper we assume that $S$ is characterized by a set of predicate symbols
(that is, if a predicate symbol $p$ appears in $S$ then every ground atom
with predicate symbol $p$ appears in $S$),
and that all the predicate symbols of the heads of rules of $R$
also occur in $S$.
A {\em program} is a module such that
every predicate which occurs in $R$ also occurs in $S$.
In what follows we will generally use $P$ also to refer to the set of rules $R$
and $def(P)$ to refer to $S$.
 
We write $P >> Q$ if no predicate of $Q$ depends on a predicate of $P$\footnote{
In \cite{stable}, this notation is used for the same idea,
but based on sets of ground atoms, rather than sets of predicates.
}.
That is, every predicate which appears in the body of a rule of $Q$
does not appear in $def(P)$.
This includes the possibility that $Q$ contains only unit rules.
We can view this as saying that the module $P$ might call the module $Q$,
but never vice versa.

One simple example of $P >>Q$ occurs when we wish to extend a statement
about atoms which is directly expressible in terms of, say, $lfp(f_Q)$
to a statement about goals.
One technique that often works is to consider the program
$P \cup Q$ where $P$ is
$\{Answer(\tilde{x}) \leftarrow Goal\}$
and $\tilde{x} = vars(Goal)$.
For example,
$\{ Goal\theta \} \subseteq lfp(f_P)$ iff
$Answer(\tilde{x})\theta \in lfp(f_{P \cup Q})$.
It is clear that $P >>Q$ in this case.
 
\section{The Lemma}
 
A complete partial order is a partially ordered set $(S, \leq)$ with a least element
where the least upper bound of a chain of elements always exists.
That is, for any $X_1 \leq \cdots X_j \leq \cdots$, $\sqcup_i X_i$ is defined.
Every complete lattice is a complete partial order.

Let $f$ and $g$ be functions on a complete partial order.
We define $(f+g)(X) = f(X) \sqcup g(X)$,
$f^+(X) = f(X) \sqcup X$,
$f^\alpha$ denotes $f$ applied $\alpha$ times in the usual way
($\alpha$ may be transfinite),
$f \leq g$ iff for every $X$, $f(X) \leq g(X)$.
$f$ is monotonic if $X \leq Y$ implies $f(X) \leq f(Y)$, for every $X$ and $Y$.
$f$ is increasing if $f(X) \geq X$ for every $X$.
Thus, for any function $f$, $f^+$ is the smallest increasing function
which is greater than $f$.
$X$ is a fixedpoint of $f$ if $f(X) = X$.
Every fixedpoint of $f$ is also a fixedpoint of $f^+$.
 
$f^*$ is the operation of closing under $f$, that is,
$f^*(X) = \sqcup_{\beta \leq \alpha} f^\beta(X)$
for some, possibly transfinite, ordinal $\alpha$.
We stipulate that $f^0(X) = X$.
If $f$ is monotonic and $X \leq f(X)$ (in particular, if $f$ is increasing)
then
$f^*(X)$ is the least fixedpoint of $f$ greater than $X$,
which we also denote by $lfp(f, X)$.
We also have $f^*(X) = lfp(f^+, X)$.

Clearly, $f \leq f^+ \leq f^*$.
Furthermore,  
$f^* = f \comp f^*$, $f^* = f^* + f^* = f^* \comp f^*$ and $f^* = (f^*)^*$.

All expressions involving $+$, $\comp$, $^*$ and monotonic, increasing functions
also represent monotonic, increasing functions.
Also, for any expression $e_1$ and $e_2$, if $e_1 \leq e_2$ then both
$e \comp e_1 \leq e \comp e_2$ and $e_1 \comp e \leq e_2 \comp e$, for any expression $e$.

The operator $^*$ is monotonic on monotonic functions.
That is $f \leq g$ implies $f^* \leq g^*$.
Proof is by transfinite induction.
Let $X$ be an element of the domain.
$f^0(X) = X = g^0(X)$.
For successor ordinals $\beta+1$,
$f^{\beta+1}(X) = f( f^\beta(X) ) \leq f( g^\beta(X) ) \leq g( g^\beta(X) ) = g^{\beta+1}(X)$,
using monotonicity of $f$ and $f \leq g$.
For limit ordinals $\alpha$,
$f^{\alpha}(X) = \sqcup_{\beta < \alpha} f^{\beta}(X) \leq \sqcup_{\beta < \alpha} g^{\beta}(X) = g^{\alpha}(X)$.
Thus $^*$ is a closure operator on monotonic functions.

\begin{lemma}
Let $f$ and $g$ be monotonic, increasing functions on a complete partial order ordered by $\geq$.
\begin{enumerate}
\item
$(f+g)^* = (f \comp g)^*  = (g \comp f)^* $.

\item
If $f^* \comp g \geq g \comp f^*$ then
$(f+g)^* = (f \comp g)^* = f^* \comp g^*$.
 
\item
If $g$ is continuous and $f \comp g \geq g \comp f$ then
$(f+g)^* = (f \comp g)^* = f^* \comp g^*$.

\item
If $g$ is continuous and $f \comp g^* \geq g^* \comp f$ then
$(f+g)^* = (f \comp g)^* = f^* \comp g^*$.

\end{enumerate}
\end{lemma}
\begin{proof}
Part 1.
$(f+g) \leq (f \comp g)$, so $(f+g)^* \leq (f \comp g)^*$.
$ (f \comp g) \leq (f+g)^2$, so $(f \comp g)^* \leq  (f+g)^*$.
Thus $(f+g)^* = (f \comp g)^*$.
By symmetry, we also have $(f+g)^* = (g \comp f)^*$.

Part 2.
$(f+g)^* = (f+g)^* \comp (f+g)^* \geq  f^* \comp g^*$,
using $(f+g) \geq f$ and $(f+g) \geq g$.
This inequality holds without the need for the hypothesis.

For the other inequality, observe that
$g \comp f^* \comp g^* \leq f^* \comp g \comp g^* = f^* \comp g^*$.
Hence
 
$(f+g) \comp f^* \comp g^*  \\
= f \comp f^* \comp g^*  + g \comp f^* \comp g^* \\
=         f^* \comp g^*  + g \comp f^* \comp g^* \\
=         f^* \comp g^*$
 
using the above observation.
Since, $f^* \comp g^*(X)$ is closed under $(f+g)$,
we must have
$(f+g)^* \leq f^* \comp g^*$.

Part 3.
We show that the hypotheses imply the hypothesis of part 1.
We claim $g \comp f^\beta \leq f^\beta \comp g$ for every $\beta$.
The proof is by transfinite induction.
For $\beta = 0$, it reduces to $g \leq g$.
For a successor ordinal $\beta+1$, we have
$g \comp f^{\beta+1} = g \comp f \comp f^{\beta} 
\leq  f \comp g \comp f^{\beta} 
\leq  f \comp f^{\beta} \comp g = f^{\beta+1}  \comp g$,
using the second hypothesis of part 3.

For a limit ordinal $\alpha$, \\
$g \comp f^\alpha \\
= g \comp ( \sqcup_{\beta < \alpha} ~ f^\beta ) \\
= ( \sqcup_{\beta < \alpha} ~ g \comp f^\beta ) \\
\leq ( \sqcup_{\beta < \alpha} ~ f^\beta \comp g) \\
= ( \sqcup_{\beta < \alpha} ~ f^\beta ) \comp g \\
= f^\alpha  \comp g $. \\
In this derivation we use both hypotheses of part 3.

Since $f^*$ is $f^\alpha$, for some ordinal $\alpha$,
$g \comp f^* \leq f^*  \comp g $.
Now, by part 2, $(f+g)^* = f^* \comp g^*$.

Part 4.
We can apply Part 3 with $g^*$ in place of $g$, provided we can prove that $(f+g)^* = (f+g^*)^*$,
and that $g^*$ is continuous.
Now,
$(f+g)^* = ((f+g)^*)^* = ((f+g)^* + (f+g)^*)^* \geq (f + g^*)^* \geq (f+g)^*$,
using monotonicity and straightforward inequalities.
Hence, $(f+g)^* = (f+g^*)^*$.

Let $\{X_i\}$ be an increasing sequence of elements of the complete partial order.
We prove $g^\alpha(\sqcup_i X_i) = \sqcup_i g^\alpha(X_i)$, for every ordinal $\alpha$, by transfinite induction.
Then $g(\sqcup_i X_i) = \sqcup_i g(X_i)$, since $g$ is continuous.
For a successor ordinal $\beta+1$ we have
$g^{\beta+1}(\sqcup_i X_i) = g \comp g^{\beta}(\sqcup_i X_i) = g( \sqcup_i g^{\beta}(X_i) ) = \sqcup_i g^{\beta+1}(X_i)$.
For a limit ordinal $\alpha$ we have
$g^\alpha(\sqcup_i X_i) = \sqcup_{\beta < \alpha} g^{\beta}(\sqcup_i X_i) 
= \sqcup_{\beta < \alpha} \sqcup_i g^{\beta}(X_i) 
=  \sqcup_i \sqcup_{\beta < \alpha} g^{\beta}(X_i) 
=  \sqcup_i g^\alpha(X_i)$.
Since $g^* = g^\alpha$, for some $\alpha$, $g^*$ is continuous.
\end{proof}
 
\section{Application to logic programming semantics}

The usefulness of this lemma comes from the following observations:
\begin{itemize}
\item
if $f$ is continuous then so is $f^+$;
if $f$ is monotonic then so is $f^+$;
\item
if $f$ is monotonic and $lfp(f, X)$ exists then
$f^*(X) = lfp(f, X) = lfp(f^+, X) = f^{+*}(X)$;
\item
in the context of logic programs, there are many semantics defined as the
least fixedpoint of a monotonic function $f_P$,
dependent on a program $P$.
In the context of modules, the semantics becomes the closure,
under $f_P$, of the semantics of the modules on which $P$ depends.
Furthermore, we often have
\begin{itemize}
\item
$f_{P \cup Q} = f_P + f_Q$,
or the weaker
$f_P + f_Q \leq f_{P \cup Q} \leq f_P \comp f_Q$
(perhaps provided that $P >> Q$).
It is straightforward to show that
$(f+g)^* = (f \comp g)^* = (g \comp f)^*$
when $f$ and $g$ are monotonic and increasing.
Thus in these cases we have
$f_{P \cup Q}^* = (f_P + f_Q)^*$.
\item
if $P >> Q$ then
$f_P^* \comp f_Q \geq f_Q \comp f_P^*$
and/or
$f_P \comp f_Q \geq f_Q \comp f_P$.
In fact, we often have
$f_P + f_Q = f_Q \comp f_P$, which can be easy to show.
If $P >> Q$ does not hold then generally these properties do not hold,
although they do if $g = c^+$ for some constant function $c$
and in some other cases (see, for example, \cite{lm,thesis}).
\end{itemize}
\end{itemize}
Applying the lemma in these cases, we have that
$(f_{P \cup Q})^* = f_P^* \comp f_Q^*$,
so the structure of the closure semantics reflects
the modular structure of the program $P \cup Q$.
 
For example, using $T_P$ \cite{vek} for definite programs $P$,
$T_P^*$ is $\lsb P \rsb$ \cite{lm} and we have:
if $P >> Q$
then $\lsb P \cup Q \rsb = \lsb P \rsb \comp \lsb Q \rsb$.
(We take $f = T_P^+ = T_P + Id$ and $g = T_Q^+ = T_Q + Id$,
where $Id$ is the identity function.)
In terms of fixedpoints, we get:
if $P >> Q$ then 
$lfp(T_{P \cup Q}, X) = lfp(T_P, lfp(T_Q, X))$,
provided $lfp(T_{P \cup Q}, X)$ exists.
Just as quickly we get similar results for
semantics based on sets of atoms possibly containing variables \cite{flmp},
and sets of clauses (possibly enhanced) \cite{bm,bglm,bcgm},
even when the function encodes a left-to-right selection rule.
Equally, the technique applies to
constraint logic programs \cite{jl},
programs involving universal quantification
\cite{plaza},
and weighted programs such as \cite{Shapiro,vs}.

The main points of this discussion are summarized in the following
proposition.
\begin{prop}
Let $P$ and $Q$ be programs, and $f$ a semantic function over a complete partial order ordered by $\geq$.
If
\vspace{-0.4cm} 
\begin{enumerate}
\item
$f_P$ and $f_Q$ are monotonic,
\item
$f_P^+ + f_Q^+ \leq f_{P \cup Q}^+ \leq f_P^+ \comp f_Q^+$
\item
Either
\begin{itemize}
\item
$f_Q$ is continuous,
and either
$f_Q^+ \comp f_P^+ \leq f_P^+ \comp f_Q^+$ or
$f_Q^* \comp f_P^+ \leq f_P^+ \comp f_Q^*$, or
\item
$f_Q^+ \comp f_P^* \leq f_P^* \comp f_Q^+$
\end{itemize}
\hspace*{-\labelsep} \hspace*{-\labelwidth}
then
$f_{P \cup Q}^* = f_P^* \comp f_Q^*$. 

\vspace{0.2cm} 
\hspace*{-\labelsep} \hspace*{-\labelwidth} 
If, further,
\item
$f_{P \cup Q}$ (or $f_P + f_Q$, or $f_P \comp f_Q$)
has a fixedpoint greater than (or equal to) $X$
\end{enumerate}
then
$lfp(f_{P \cup Q}, X) = lfp(f_P, lfp(f_Q, X))$.
\end{prop}
\begin{proof}
As noted earlier, the first two hypotheses imply that
$f_{P \cup Q}^* = (f_P + f_Q)^*$.
By the previous lemma,
$(f_P + f_Q)^* = f_P^* \comp f_Q^*$.
Thus
$f_{P \cup Q}^* = f_P^* \comp f_Q^*$,
that is,
$lfp(f_{P \cup Q}^+, X) = lfp(f_P^+, lfp(f_Q^+, X))$,
for every $X$.
It is known that if a monotonic function has a fixedpoint
greater than (or equal to) $X$,
then it has a least such fixedpoint.
It is straightforward to show that
$lfp(f_P + f_Q, X) = lfp(f_{P \cup Q}, X) = lfp(f_P \comp f_Q, X)$
if one of these exists, in which case they all do.
Since $lfp(f_{P \cup Q}, X)$ exists, so also must
$lfp(f_P + f_Q, X)$,
$lfp(f_Q, X)$ and $lfp(f_P, lfp(f_Q, X))$.
As observed above, since the fixedpoints exists,
they are equal to the fixedpoints of the corresponding increasing function,
which completes the proof.
\end{proof}
 
Note that we can weaken the second hypothesis
by replacing
$f_P^+ \comp f_Q^+$
by any expression involving
$f_P^+$, $f_Q^+$, $+$ and $\comp$
and the proposition will still hold.

The requirements of hypothesis 3 can often be tested syntactically.
For example,
for definite logic programs and the function $T_P$,
the condition
$f_Q^+ \comp f_P^+ \leq f_P^+ \comp f_Q^+$
can be tested using unfolding and subsumption \cite{thesis,equivs}.
In the case of Datalog programs, again using $T_P$, more tests are possible.
For example, sometimes it is possible to test
$f_Q^+ \comp f_P^+ \leq f_P^+ \comp f_Q^*$ \cite{RSUV}.
 
If we are simply interested in the combinations of functions
(even if $f_P$ and $f_Q$ are not associated with programs),
and not interested in $f_{P \cup Q}$,
then we have the following corollary of the above proof.
 
\begin{cor}
Under hypotheses 1 and 3 of the above proposition,
$(f_P + f_Q)^* = (f_P \comp f_Q)^* = f_P^* \comp f_Q^*$.
 
Under hypotheses 1, 3 and 4 of the above proposition,
$lfp(f_P + f_Q, X) = lfp(f_P \comp f_Q, X) = lfp(f_P, lfp(f_Q, X))$.
\end{cor}

If we have multiple modules, we might want to look at semantics in terms of common fixedpoints
and/or chaotic iterations \cite{cc,lm,thesis}.
For increasing functions $f$ and $g$, the common fixedpoints of $f$ and $g$
are exactly the fixedpoints of $f \comp g$
(or $g \comp f$, or $f + g$, or any other composition of $f$'s and $g$'s).
Formulating semantics in terms of $f^+_P$ makes the relationship with common fixedpoints
and chaotic iterations easier.
 
\subsection{Duality}
There are also the dual results to the above,
involving decreasing functions, downwards closures, greatest fixedpoints, etc.,
but these seem less useful since the dual of function addition\footnote{
The dual of function addition is $(f \ov{+} g)(X) = f(X) \sqcap g(X)$.}
occurs less often in practice,
at least in the context we consider here.
 
At the very least we have:
if $f$ and $g$ are monotonic decreasing functions
and $f^\bullet \comp g \leq g \comp f^\bullet$
then $(f \comp g)^\bullet = (g \comp f)^\bullet
= f^\bullet \comp g^\bullet $.
Here $f^\bullet$ denotes the downward closure of $f$.
But, in general,
$(f + g)^\bullet \neq f^\bullet \comp g^\bullet $
when these conditions apply.
For example,
let $\cC$ be the lattice of subsets of $\{a,b,c\}$.
Define {\sl $f(x) = \{a,b\}$ if $x=\{a,b,c\}$ and $\emptyset$ otherwise}, and
define {\sl $g(x) = \{b,c\}$ if $x=\{a,b,c\}$ and $\emptyset$ otherwise}.
Clearly $f$ and $g$ are monotonic and decreasing
and satisfy $f^\bullet \comp g \leq g \comp f^\bullet$.
However
$(f + g)^\bullet = \{a,b,c\} \neq \emptyset = f^\bullet \comp g^\bullet $.
 
If we can express
$f_{P \cup Q}$ as a functional expression of $f_P$ and $f_Q$ involving only
function composition
-- for example, $f_{P \cup Q} = f_P \comp f_Q$ --
or, more generally, bound $f_{P \cup Q}$ between two such expressions
then we do get something.
 
\section{Bottom-up evaluation}

Suppose the elements of the complete partial order can be viewed as
(possibly infinite) programs that are their own semantics.
That is, suppose that there is a mapping $m$
which maps every $X$ in the complete partial order to a program $P_X$ such that
$lfp(f_{P_X}) = X$.
Such a mapping represents an evaluation of the program.
In the following proposition we make the extra assumption
that $f_{P_X}$ is a constant function, that is,
$\forall Y ~ f_{P_X}(Y) = X$.
 
\begin{prop}
Let $P$ and $Q$ be programs, and $f$ a semantic function
on a complete partial order ordered by $\geq$.
If
\begin{enumerate}
\item
for all $P'$,
$f_{P'}$ is monotonic,
\item
for all programs ${P'}$ and ${Q'}$,
$f_{P'}^+ + f_{Q'}^+ \leq f_{{P'} \cup {Q'}}^+ \leq f_{P'}^+ \comp f_{Q'}^+$
\item
Either
\begin{itemize}
\item
$f_Q$ is continuous,
and
$f_Q^+ \comp f_P^+ \leq f_P^+ \comp f_Q^+$, or
\item
$f_Q^+ \comp f_P^* \leq f_P^* \comp f_Q^+$
\end{itemize}
\item
$f_{P \cup Q}$ (or $f_P + f_Q$, or $f_P \comp f_Q$)
has a fixedpoint greater than (or equal to) $X$
\end{enumerate}
then
$lfp(f_{P \cup Q}, X)
= lfp(f_{P \cup QX}, X)
= lfp(f_{P \cup QX})$,
where $QX$ is the program corresponding to $lfp(f_Q, X)$,
and $f_{QX}$ is the constant function 
$f_{QX}(Y) = lfp(f_Q, X)$.
\end{prop}
\begin{proof}
We apply the previous proposition and reason
 
$lfp(f_{P \cup Q}, X) \\
= lfp(f_P, lfp(f_Q, X)) \\
= lfp(f_P, lfp(f_{QX}, X)) \\
= lfp(f_{P \cup QX}, X)$
 
The last step is a second application of the previous proposition
(involving $P$ and $QX$) and needs some argument.
Since $f_{QX}$ is a constant, the third hypothesis of the proposition
is satisfied.
Let $Z = lfp(f_P, lfp(f_{QX}, X))$.
Clearly $f_P(Z) = Z$, 
and $f_{QX}(Z) \leq Z$ since $f_{QX}(Z) = lfp(f_Q, X) \leq Z$,
so that $f_P + f_{QX}$ has a fixedpoint $Z$ greater than $X$.
 
The last equality in the statement of the proposition holds
since it is clear that any fixedpoint of
$f_{P \cup QX}$
is greater than $X$.
\end{proof}
 
This proposition justifies a simple form of partial evaluation
in which $Q$ is partially evaluated wrt $X$,
and the result is added to the program in the form of $QX$.
Taking the example of definite logic programs and $T_P$,
we can take $m$ to be the identity function,
so that
$QX = lfp(T_Q, X)$ and
$lfp(T_{P \cup Q}, X) = lfp(T_{P \cup QX}, X) = lfp(T_{P \cup QX})$.
 
In many cases of interest we can weaken the condition that
$f_{QX}$ be a constant function.
If, instead, for every $P$ and $Q$,
$P >> Q$ implies hypothesis 3 is satisfied,
and we further assume that
$P >> Q$ and $P >> m(X)$ implies $P >> QX$,
then the conclusion of the above proposition holds,
even if $f_{QX}$ is not a constant.
 
%
\section{Programs with Negation}
 
The semantics of programs with negation generally make the implicit assumption
that any predicate not defined in the program has empty extension
(i.e., is false).
To handle modules, these semantics must be modified slightly,
by taking $def(P)$ into account,
so that a predicate intended to be defined in another module
is not automatically given an empty extension in the semantics of $P$.
Such a modification generally does not affect such properties as
monotonicity and continuity of the function involved.
 
Roughly speaking, we will be replacing a function $f_P$
by $f_P'$ where
$f_P'(I) = f_P(I)|_{def(P)} \sqcup I|_{\ov{def(P)}}$
where $\ov{def(P)}$ denotes the complement of $def(P)$,
and $X|_Y$ means something like $X \cap Y$.
That is, the effects of the application of $f_P$ are restricted to $def(P)$.
 
We examine Fitting's semantics \cite{fit} first.
A partial interpretation $I$ over the domain of computation
is represented by the consistent set of ground literals
which are consequences of $I$.
The function $\Phi_P$ is modified to handle modules as follows.
$\Phi_P(I) = \{ A \ |\ A \leftarrow L_1, \ldots , L_k \in gd(P),
A \in def(P),
I \models L_i, i=1,\ldots,k \}
\cup \{ \neg A \ |\ \mbox{for every } A \leftarrow L_1, \ldots , L_k \in gd(P),
A \in def(P),
I \models \neg ( L_1 \wedge \ldots \wedge L_k) \}$.
It is straightforward to see that, if $P >> Q$,
$\Phi_{P \cup Q} = \Phi_P + \Phi_Q$
and
$\Phi_Q^+ \comp \Phi_P^+ = \Phi_Q^+ + \Phi_P^+$.
Here $m_{F}(I)
= \{ A : I \models A \} \cup
  \{ A \leftarrow      A : I \not\models A, I \not\models \neg A \}$
and we have
if $P >> Q$ then
$lfp(\Phi_{P \cup Q}, X)
= lfp(\Phi_P, lfp(\Phi_Q, X))
= lfp(\Phi_{P \cup Q'}, X)$,
where $Q' = m_{F}(lfp(\Phi_Q))$.
 
We now turn our attention to the well-founded semantics \cite{grs}.
We define $\neg S = \{ \neg s \ |\ s \in S \}$
and identify $\neg \neg s$ with $s$.
Let
$T_P(I) = \{ A \ |\ A \leftarrow L_1, \ldots , L_k \in gd(P),
A \in def(P),
I \models L_i, i=1,\ldots,k \}$.
A $P,I$-unfounded set is a set $U \subseteq def(P)$ of atoms $A$ such that
for every rule $A \leftarrow L_1, \ldots , L_k$ in $gd(P)$
there is some $i$ such that either $I \models \neg L_i$ or $L_i \in U$.
Let $U_P(I)$ denote the greatest $P,I$-unfounded set
and let $W_P(I) = T_P(I) \cup \neg U_P(I)$.
Assume $J$ is a partial interpretation
defining only predicate symbols not in $def(P)$.
The least
(under the definedness ordering)
fixedpoint of $W_P$ which is greater than $J$ is a partial model of $P$,
called the {\em well-founded} partial model of $P$ extending $J$,
denoted $WF(P, J)$.
For $J = \emptyset$ we write $WF(P)$.
If $def(P)$ is the ``Herbrand base'' then
this definition reduces to the usual
definition of the well-founded partial model
\cite{grs}.
 
If $P$ is $\{ p \leftarrow p; p \leftarrow q \}$ and
$Q$ is $\{ q \leftarrow q \}$,
where $def(P) = \{ p \}$ and $def(Q) = \{ q \}$,
then $P >> Q$.
We have $W_P( \emptyset ) = \emptyset$ and
$W_Q( \emptyset ) = \{ \neg q \}$.
On the other hand,
$W_{P \cup Q}( \emptyset ) = \{ \neg p, \neg q \}$
and thus
$W_{P \cup Q}^+ \neq W_P^+ + W_Q^+$.
Nevertheless we are still able to satisfy hypothesis 2 of Proposition 4,
as we now show.
 
It can be verified that
$U_P(I \cup J) \supseteq U_P(I)$
whenever $(J \cup \neg J) \cap def(P) = \emptyset$.
This is used in the penultimate step below.
 
Clearly
$W_P^+(I) \cup W_Q^+(I) \leq W_{P \cup Q}^+(I)$.
If $P >> Q$ then
 
$W_{P \cup Q}^+(I) \\
= T_{P \cup Q}(I) \cup \neg U_{P \cup Q}(I) \\
= T_P(I) \cup T_Q(I) \cup I \cup \neg U_Q(I) \cup \neg U_P(I \cup \neg U_Q(I)) \\
= T_P(I) \cup W_Q^+(I) \cup \neg U_P(I \cup \neg U_Q(I)) \\
\leq T_P(W_Q^+(I)) \cup W_Q^+(I) \cup \neg U_P(W_Q^+(I)) \\
= W_P^+(W_Q^+(I))$
 
Also in this case, we have that \\
$(W_Q^+ \comp W_P^+ )(I) \\
= T_Q( W_P^+(I) ) \cup W_P^+(I) \cup \neg U_Q( W_P^+(I) ) \\
= T_Q( I ) \cup W_P^+(I) \cup \neg U_Q( I ) \\
= (W_Q^+ + W_P^+ )(I) \\
\leq (W_P^+ \comp W_Q^+ )(I)$
 
Since $W_P$ is known to be monotonic \cite{grs},
applying the above proposition gives us:
if $P >> Q$ then
$WF(P \cup Q, X) = WF(P, WF(Q, X)) = WF(P \cup Q', X)$
where $m_{WF}(I)
= \{ A : A \in I \} \cup
  \{ A \leftarrow      A : \neg A \in I \} \cup
  \{ A \leftarrow \neg A : A \not\in I, \neg A \not\in I \}$,
and $Q' = m_{WF}(WF(Q))$.
In particular, if $P$ depends only on $Q$ and $Q$ is a program
(i.e. only depends on itself) then
$WF(P \cup Q) = WF(P \cup m_{WF}(WF(Q)))$.
 
\section{Conclusion}

Closure operators are the natural semantics for modules when semantics is defined by least fixedpoints. 
Working with monotonic, increasing functions is more convenient than simply monotonic functions.
Passing from $f$ to $f^+$ makes reasoning easier.
 
Of course, this technique is dependent on an appropriate least fixedpoint
characterization of the semantics.
Thus it is not directly applicable to
the Clark-completion semantics \cite{clark},
Kunen's semantics \cite{Kun87},
the stable model \cite{GL88}
and stable class semantics \cite{bs}.
But perhaps Fage's semantics.....
\cite{Fag89}

\textbf{Acknowledgement}
This work was conducted while the author was an employee of IBM.

\appendix

\newcommand{\FPT}{{\mathcal{FPT}}}
\newcommand{\PRE}{{\mathcal{PRE}}}
\newcommand{\POST}{{\mathcal{POST}}}

\section{Fixedpoints of a Sandwiched Function}

We review some notions of fixedpoints of functions on a partially ordered set $(S, \leq)$.
$X$ is a \emph{pre-fixedpoint} of a function $f$ if $f(X) \leq X$;
thus a pre-fixedpoint is closed under the action of $f$.
$X$ is a \emph{fixedpoint} of $f$ if $f(X) = X$.
$X$ is a \emph{post-fixedpoint} of $f$ if $X \leq f(X) $.
Let $\PRE(f)$, $\POST(f)$ and $\FPT(f)$ denote, respectively, the set of pre-fixedpoints, post-fixedpoints, and
fixedpoints of a function $f$.
Note that every fixedpoint is also a pre-fixedpoint and a post-fixedpoint.

The following lemma shows that a function that is intermediate between two functions with the same
pre-fixedpoints and fixedpoints, has exactly the same pre-fixedpoints and fixedpoints as those functions.
Note that this lemma does not require the functions to be monotonic, 
nor does it require any conditions on the partial order.

\begin{lemma}
Let $f_1$, $f_2$ and $g$ be functions on a partially ordered set $(S, \leq)$.
Suppose for every $X$, $f_1(X) \leq g(X) \leq f_2(X)$.

\begin{enumerate}
\item
If $\PRE(f_1) = \PRE(f_2)$
then
$\PRE(g) = \PRE(f_1) = \PRE(f_2)$
\item
If $\POST(f_1) = \POST(f_2)$
then
$\POST(g) = \POST(f_1) = \POST(f_2)$
\item
If $\PRE(f_1) = \PRE(f_2)$, and
$\FPT(f_1) = \FPT(f_2)$
then
$\FPT(g) = \FPT(f_1) = \FPT(f_2)$
\item
If $\POST(f_1) = \POST(f_2)$, and
$\FPT(f_1) = \FPT(f_2)$
then
$\FPT(g) = \FPT(f_1) = \FPT(f_2)$
\item
If $\PRE(f_1) = \PRE(f_2)$, and
$\POST(f_1) = \POST(f_2)$ then
$\FPT(g) = \FPT(f_1) = \FPT(f_2)$
\item
If $\FPT(f_1) = \FPT(f_2)$ then $\FPT(g) \supseteq \FPT(f_1) = \FPT(f_2)$
\end{enumerate}
\end{lemma}
\begin{proof}
Part 1.
Suppose $\PRE(f_1) = \PRE(f_2)$.
Now, let $I$ be a pre-fixedpoint of $g$.
Then $f_1(I) \leq g(I) \leq I$.
Thus $I$ is a pre-fixedpoint of $f_1$ and $f_2$.
Conversely, let $I$ be a pre-fixedpoint of $f_1$ and $f_2$.
Then $g(I) \leq f_2(I) \leq I$.
Hence $I$ is a pre-fixedpoint of $g$.
Thus $\PRE(g) = \PRE(f_1) = \PRE(f_2)$.

Part 2.
By duality, the first part implies:
If $\POST(f_1) = \POST(f_2)$
then
$\POST(g) = \POST(f_1) = \POST(f_2)$.

Part 3.
Suppose $\PRE(f_1) = \PRE(f_2)$, and $\FPT(f_1) = \FPT(f_2)$.
Let $I$ be a fixedpoint of $f_1$ and $f_2$.
Then $I = f_1(I) \leq g(I) \leq f_2(I) =I$.
Hence $g(I) = I$.
Conversely, let $I$ be a fixedpoint of $g$.
By Part 1, $\PRE(g) = \PRE(f_2)$ and hence $f_2(I) \leq I$.
Also, $I = g(I) \leq f_2(I)$.
Thus $I$ is also a fixedpoint of $f_2$ and, hence,
$\FPT(g) = \FPT(f_1) = \FPT(f_2)$.

Part 4.
This is the dual of part 3.

Part 5.
If $\PRE(f_1) = \PRE(f_2)$, and
$\POST(f_1) = \POST(f_2)$ then, by parts 1 and 2,
$\PRE(g) = \PRE(f_1) = \PRE(f_2)$ and $\POST(g) = \POST(f_1) = \POST(f_2)$.
The fixedpoints are those elements that are both a pre- and post-fixedpoint.
Hence, $\FPT(g) = \FPT(f_1) = \FPT(f_2)$.

Part 6.
Suppose $\FPT(f_1) = \FPT(f_2)$,
and let $I \in \FPT(f_1)$.
Then $I = f_1(I) \leq g(I) \leq f_2(I) = I$,
so $I$ is also a fixedpoint of $g$.
Hence $\FPT(f_1) = \FPT(f_2) \subseteq \FPT(g)$.
\end{proof}

It is tempting to assume that we could have:
if $\FPT(f_1) = \FPT(f_2)$ then $\FPT(g) = \FPT(f_1) = \FPT(f_2)$.
However, this does not hold, in general, as the following example shows.

\begin{example}
Let $S = \{1, 2, 3 \}$ under the usual ordering.
We define $f_1$ and $f_2$ as follows:
$f_1(1) = f_1(2) = 1$ and $f_1(3) = 3$.
$f_2(1) = 1$ and $f_2(2) = f_2(3)=3$.
Let $g$ be the identity function.
It is straightforward to verify that
$f_1$, $f_2$ and $g$ are monotonic functions, $\FPT(f_1) = \FPT(f_2) = \{1, 3\}$,
and for all $X$, $f_1(X) \leq g(X) \leq f_2(X)$.
However, 2 is a fixedpoint of $g$, but not of $f_1$ or $f_2$.
\end{example}


\begin{thebibliography} {99}
\renewcommand{\ignore}[1]{}
\small
 
\bibitem{bs}
C. Baral \& V.S. Subrahmanian,
Stable and Extension Class Theory for Logic Programs and Default Logics,
{\em Journal of Automated Reasoning} 8 (3), 345--366, 1992.
 
\bibitem{bcgl}
R. Barbuti, M. Codish, R. Giacobazzi \& G. Levi,
Modelling Prolog Control,
{\em Proc. POPL}, 95--104, 1992.
 
\bibitem{bcgm}
R. Barbuti, M. Codish, R. Giacobazzi \& M. Maher,
Oracle Semantics for Prolog,
{\em Proc. 3rd Int. Conf.  Algebraic and Logic Programming}, 100--114, 1992.
 
\bibitem{bm}
A. Bossi \& M. Menegus,
Una Semantica Composizionale per Programmi Logici Aperti,
{\em Proc. 6th Italian Conf. on Logic Programming}, 95--100, 1991.
 
\bibitem{bglm}
A. Bossi, M. Gabbrielli, G. Levi \& M.C. Meo,
Contributions to the Semantics of Open Logic Programs,
{\em Proc. Int. Conf. on Fifth Generation Computer Systems},
570--580, 1992.
 
\bibitem{clark}
K. Clark,
Negation as Failure,
in: {\em Logic and Databases},
H. Gallaire \& J. Minker (Eds),
Plenum Press, 293-322, 1978.

\bibitem{cc}
P. Cousot \& R. Cousot,
Constructive versions of Tarski's fixed point theorems,
{\em Pacific J. Math.} 82,  1 (1979), 43--57.

\bibitem{vek}
M.H. van Emden \& R.A. Kowalski,
The Semantics of Predicate Logic as a Programming Language,
{\em Journal of the ACM 23}, 4 (1976), 733--742.
 
\bibitem{Fag89}
F. Fages,
A New Fixpoint Semantics for General Logic Programs
Compared with the Well-Founded and the Stable Model Semantics,
{\em Proc. ICLP-7}, 441--458, 1990.
 
\bibitem{flmp}
M. Falaschi, G. Levi, M. Martelli \& C. Palamidessi,
Declarative Modeling of the Operational Behavior of Logic Languages,
{\em Theoretical Computer Science}, 69, 289--318, 1989.
 
\bibitem{fit}
M. Fitting,
A Kripke-Kleene Semantics for Logic Programs,
{\em Journal of Logic Programming}, 4, 295--312, 1985.
 
\bibitem{fit2}
M. Fitting,
Well-founded Semantics, Generalized,
{\em Proc. ILPS},
71--84, 1991.
 
 
\bibitem{grs}
A. van Gelder, K. Ross \& J.S. Schlipf,
Unfounded Sets and Well-Founded Semantics for General Logic Programs,
{\em Proc. PODS'88}, 221--230, 1988.
 
\bibitem{GL88}
M. Gelfond \& V. Lifschitz,
The Stable Model Semantics for Logic Programming,
{\em Proc. ICLP/SLP-5}, 1070--1080, 1988.

\bibitem{IW91}
Y.E. Ioannidis \& E. Wong,
Towards an Algebraic Theory of Recursion, {\em JACM} 38(2), 329--381, 1991.
 
\bibitem{jl}
J. Jaffar \& J-L. Lassez,
Constraint Logic Programming,
{\em Proc. POPL},
111--119, 1987.
 
\bibitem{KS90}
K. Kanchanasut \& P. Stuckey,
Eliminating Negation from Normal Logic Programs,
{\em Proc. ALP'90}, 217--231,
1990. 
 
\bibitem{Kun87}
K. Kunen,
Negation in Logic Programming,
{\em Journal of Logic Programming}, 4, 289--308, 1987.
 
\bibitem{Kun89}
K. Kunen,
Signed Data Dependencies in Logic Programs,
{\em Journal of Logic Programming}, 7, 231--245, 1989.
 
\bibitem{lm}
J-L. Lassez \& M.J. Maher,
Closures and Fairness in the Semantics of Logic Programs,
{\em Theoretical Computer Science}, 29, 167--184, 1984.
 
\bibitem{thesis}
M.J. Maher,
Semantics of Logic Programs,
Ph.D. thesis,
Technical Report TR85/14,
Department of Computer Science, University of Melbourne, 1985.
 
\bibitem{equivs}
M.J. Maher,
Equivalences of Logic Programs,
in: {\em Foundations of Deductive Databases and Logic Programming},
J. Minker (Ed), Morgan-Kaufmann, 627--658, 1988.

\bibitem{modules}
M.J. Maher,
A Transformation System for Deductive Database Modules with Perfect Model Semantics, 
{\em Proc. FSTTCS}, 89--98, 1989.

\bibitem{stable}
M.J. Maher,
Reasoning about Stable Models (and other Unstable Semantics),
manuscript, 1990.
 
\bibitem{plaza}
J. Plaza,
Fully Declarative Programming with Logic: Mathematical Foundations,
Ph.D. thesis, City University of New York, 1990.
 
\bibitem{PP89}
H. Przymusinska \& T. Przymusinski,
Semantic Issues in Deductive Databases and Logic Programs,
in:
{\em  Formal Techniques in Artificial Intelligence},
A. Banerji (Ed.), North-Holland, 321--367, 1990.
 
\bibitem{RSUV}
R. Ramakrishnan, Y. Sagiv, J. Ullman \& M. Vardi,
Proof-tree Transformation Theorems and their Applications,
{\em Proc. PODS}, 172--181, 1989.
 
\bibitem{Shapiro}
 E.Y. Shapiro,
Logic Programs With Uncertainties: A Tool for Implementing Rule-Based Systems, 
{\em Proc. IJCAI}, 529--532, 1983.

\bibitem{vs}
V. S. Subrahmanian,
On the Semantics of Quantitative Logic Programs, {\em SLP},173--182, 1987.

\end{thebibliography}
\end{document}